\documentstyle[12pt,epsfig]{article}
\newcommand{\bfk}{\mbox{\boldmath $k$}}

\def\ii{\'{\i}}
\def\nostrocostruttino#1\over#2{\mathrel{\mathop{\kern 0pt \rlap
{\hbox{$#1$}}} \hbox{\kern-.135em $#2$}}}

\newcommand{\ZP}[1]{{\it Z.\ Phys.}\ {\bf #1}}
\newcommand{\PL}[1]{{\it Phys.\ Lett.}\ {\bf #1}}

\newcommand{\beq}{\begin{equation}}
\newcommand{\eeq}{\end{equation}}
\newcommand{\barr}{\begin{eqnarray}}
\newcommand{\earr}{\end{eqnarray}}
\newcommand{\ba}{\begin{array}}
\newcommand{\ea}{\end{array}}
\def\lsim{\mathrel{\rlap{\lower4pt\hbox{\hskip1pt$\sim$}}\raise1pt\hbox{$<$}}} 
\def\gsim{\mathrel{\rlap{\lower4pt\hbox{\hskip1pt$\sim$}}\raise1pt\hbox{$>$}}}

\newcommand{\qq}{q\bar q}
\newcommand{\la}{\lambda}
\begin{document}

%%%%%%%%%%%%%%%%%%%%%%%%%%%%%%%%%%%%%%%%%%%%%%%%%%%%%%%%%%%%%%%%%%%%%%%%%%%%%%
\begin{flushright}
DFTT 19/99 \\
INFNCA-TH9905 \\
hep-ph/9904205 \\
\end{flushright}
\vskip 1.5cm
\begin{center}
{\bf 
Off-diagonal helicity density matrix elements for vector mesons
produced in polarized {\mbox{\boldmath $e^+e^-$}} processes
}\\
\vskip 1.5cm
{\sf M. Anselmino$^1$, M. Bertini$^{2}$, F. Caruso$^{3,4}$, F. Murgia$^5$ and 
P. Quintairos$^{6}$}
\vskip 0.8cm
{$^1$Dipartimento di Fisica Teorica, Universit\`a di Torino and \\
      INFN, Sezione di Torino, Via P. Giuria 1, 10125 Torino, Italy\\
\vskip 0.5cm
$^2$Theoretical Physics, Lund University \\ 
S\"olvegatan 14a, S-223 62 Lund, Sweden \\
\vskip 0.5cm
$^3$Centro Brasileiro de Pesquisas F\ii sicas \\
R. Dr. Xavier Sigaud 150, 22290-180 Rio de Janeiro, Brazil \\
\vskip 0.5cm
$^4$Instituto de F\ii sica da UERJ \\ 
Rua S\~ao Francisco Xavier 524, 20559-900 Rio de Janeiro, Brazil\\
\vskip 0.5cm
$^5$INFN, Sezione di Cagliari and Dipartimento di Fisica, Universit\`a di
Cagliari \\
C.P. 170, I-09042 Monserrato (CA), Italy \\
\vskip 0.5cm
$^6$Instituto de F\ii sica Te\'orica - UNESP\\
Rua Pamplona 145, 01405-900 S\~ao Paulo, Brazil }\\
\end{center}
\vskip 1.5cm
\noindent
{\bf Abstract:}

\vspace{6pt}

\noindent
Final state $\qq$ interactions give origin to non zero values of the 
off-diagonal element $\rho_{1,-1}$ of the helicity density matrix of vector 
mesons produced in $e^+ e^-$ annihilations, as confirmed by recent OPAL 
data on $\phi$, $D^*$ and $K^*$'s. New predictions are given for $\rho_{1,-1}$
of several mesons produced at large $x_{_E}$ and small $p_T$ -- {\it i.e.}
collinear with the parent jet -- in the annihilation of polarized $e^+$
and $e^-$; the results depend strongly on the elementary dynamics and 
allow further non trivial tests of the Standard Model.

%%%%%%%%%%%%%%%%%%%%%%%%%%%%%%%%%%%%%%%%%%%%%%%%%%%%%%%%%%%%%%%%%%%%%%%%%%%%%%%
\newpage
%%%%%%%%%%%%%%%%%%%%%%%%%%%%%%%%%%%%%%%%%%%%%%%%%%%%%%%%%%%%%%%%%%%%%%%%%%%%%%%
\pagestyle{plain}
\setcounter{page}{1}
\noindent
{\bf 1 - Introduction}
\vskip 12pt

In a series of papers \cite{akp}--\cite{abmq} it was pointed out how the
final state interactions between the $q$ and $\bar q$ produced
in $e^+ e^-$ annihilations -- usually neglected, but indeed 
necessary -- might give origin to non zero values of spin observables which 
would otherwise be forced to vanish. The off-diagonal spin density
matrix element $\rho_{1,-1}(V)$ 
of vector mesons may be sizeably different from zero \cite{akp, aamr} due to a 
coherent fragmentation process which takes into account $q \bar q$ 
interactions; indeed, predictions were given \cite{abmq} for
several spin 1 particles produced at
LEP in two jet events, provided they carry a large fraction
$x_{_E}$ of the parent quark energy and have a small intrinsic $\bfk_\perp$,
{\it i.e.} they are collinear with the parent jet. 

The values of $\rho_{1,-1}(V)$ are related to the values of the off-diagonal
helicity density matrix element $\rho_{+-;-+}(\qq)$ of the $\qq$ pair,
generated in the $e^- e^+ \to \qq$ process \cite{abmq}:
\beq
\rho^{\,}_{1,-1}(V) \simeq [1 - \rho^{\,}_{0,0}(V)] \>
\rho_{+-;-+}(\qq) \label{old}
\eeq
where the value of the diagonal element $\rho^{\,}_{0,0}(V)$ can be taken from
data. The values of $\rho_{+-;-+}(\qq)$ depend on the elementary short distance
dynamics and can be computed in the Standard Model. Thus, a measurement
of $\rho_{1,-1}(V)$ is a further test of the constituent dynamics, more
significant than the usual measurement of cross-sections in that it depends
on the product of different elementary amplitudes, rather than on squared
moduli. With unpolarized $e^+$ and $e^-$
\beq
\rho^{\,}_{+-;-+}(\qq) = {1\over 4N_{\qq}} \sum_{\la^{\,}_{-}, \la^{\,}_{+}}
M^{\,}_{+-;\la^{\,}_{-} \la^{\,}_{+}} \> M^*_{-+; \la^{\,}_{-} \la^{\,}_{+}}
\,, \label{rhoun}
\eeq
where the $M$'s are the helicity amplitudes for the $e^- e^+ \to \qq$ process 
and
\beq
4N_{\qq} =
\sum_{\la^{\,}_q, \la^{\,}_{\bar q}; \la^{\,}_{-}, \la^{\,}_{+}} \vert
M^{\,}_{\la^{\,}_q \la^{\,}_{\bar q}; \la^{\,}_{-} \la^{\,}_{+}} \vert^2 \,.
\label{nqq}
\eeq

At LEP energy, $\sqrt s = M_{_Z}$, one has \cite{abmq}
\beq
\rho^{\,}_{+-;-+}(\qq) \simeq
\rho^{^Z}_{+-;-+}(\qq) \simeq {1\over 2} \> {(g^2_{_V} - g^2_{_A})_q \over
(g^2_{_V} + g^2_{_A})_q} \, {\sin^2\theta \over 1+ \cos^2\theta} \, \cdot
\label{rhozap}
\eeq
where $g_{_V}$ and $g_{_A}$ are the Standard Model coupling constants
[reported for convenience in Eq. (\ref{cc})] and $\theta$ is the vector
meson production angle in the $e^-e^+$ c.m. frame.

At lower energies, where weak interactions can be neglected, one has:
\beq
\rho^{\,}_{+-;-+}(\qq) \simeq
\rho^{^\gamma}_{+-;-+}(\qq) = {1\over 2} \> 
{\sin^2\theta \over 1+ \cos^2\theta} \, \cdot
\label{rhog}
\eeq

Eq. (\ref{old}) is in good agreement with OPAL Collaboration data on
$\phi$, $D^*$ and $K^*$, including the $\theta$ dependence induced by Eq.
(\ref{rhozap}) \cite{opal1, opal2}; however, no sizeable value of
$\rho_{1,-1}(V)$ for $V= \rho, \phi$ and $K^*$ was observed by DELPHI
Collaboration \cite{delphi}. Further tests are then necessary.
Predictions for $\rho_{1,-1}(V)$, with $V= \phi, D^*$ or $B^*$
produced in $NN \to VX$, $\gamma N \to VX$ and $\ell N \to
\ell VX$ processes were given in Ref. \cite {abmp}.

We consider here again the process $e^+ e^- \to VX$, assuming all
possible polarization states for the initial leptons. This might not be a
realistic case -- polarized $e^+ e^-$ beams might not be available
in the nearest future -- but, as we shall see, the results show such
a strong interesting dependence on the spin elementary
dynamics, that such a possibility should not be forgotten when planning
future $e^+ e^-$ colliders. Also, this work is the natural expansion 
and completion -- with all possible cases and theoretical predictions
taken into account -- of the study undertaken in Ref. \cite{abmq}. 

In the next Section we compute the value of $\rho_{+-;-+}(\qq)$ with 
the most general spin states of $e^+$ and $e^-$; in Section 3
we obtain numerical estimates in several particular cases 
and in Section 4 we give some comments and conclusions.

\vskip 12pt
\noindent
{\bf 2 - Computation of} {\mbox{\boldmath $\rho_{+-;-+}(\qq)$}}
\vskip 12pt
In case of polarized initial leptons Eq. (\ref{rhoun}) modifies into:
\beq
\rho^{pol}_{\la^{\,}_q,\la^{\,}_{\bar q};\la'_q,\la'_{\bar q}}(q\bar q)
= \frac{1}{N^{pol}_{\qq}} \,
\sum_{\la^{\,}_-,\la^{\,}_+, \la'_-,\la'_+}
\, M^{\,}_{\la^{\,}_q,\la^{\,}_{\bar q};\la^{\,}_-,\la^{\,}_+}
\> \rho^{\,}_{\la^{\,}_-,\la^{\,}_+;\la'_-,\la'_+}
\> M^*_{\la'_q,\la'_{\bar q};\la'_-,\la'_+}
\label{rhopo}
\eeq
with
\beq
N^{pol}_{\qq} = 
\sum_{\la^{\,}_q,\la^{\,}_{\bar q};\la^{\,}_-,\la^{\,}_+, \la'_-,\la'_+}
\, M^{\,}_{\la^{\,}_q,\la^{\,}_{\bar q};\la^{\,}_-,\la^{\,}_+}
\> \rho^{\,}_{\la^{\,}_-,\la^{\,}_+;\la'_-,\la'_+}
\> M^*_{\la_q,\la_{\bar q};\la'_-,\la'_+}
\label{norpo}
\eeq
and where
\beq
\rho_{\la^{\,}_-,\la^{\,}_+;\la'_-,\la'_+}(e^-e^+) = 
\rho_{\la^{\,}_-,\la'_-}(e^-)\>\rho_{\la^{\,}_+,\la'_+}(e^+)
\label{rhoee}
\eeq 
is the helicity density matrix of the incoming independent leptons. 

The most general helicity density matrices for the incoming $e^-$ and $e^+$
are given by    
\beq
\rho(e^-) = \frac{1}{2} \left( \begin{array}{cc}
1+\cos\alpha_- & e^{-i\beta_-} \sin\alpha_- \\ 
e^{i\beta_-}\sin\alpha_- & 1-\cos\alpha_-
\end{array}
\right)         
\label{rhoe-}
\eeq
and
\beq
\rho(e^+) = \frac{1}{2} \left( \begin{array}{cc}
1-\cos\alpha_+ & e^{i\beta_+} \sin\alpha_+ \\ 
e^{-i\beta_+}\sin\alpha_+ & 1+\cos\alpha_+
\end{array} \right)
\label{rhoe+}
\eeq
where $\alpha_-$ and $\beta_-$ ($\alpha_+$ and $\beta_+$) are respectively
the polar and azimuthal angle of the $e^-$ ($e^+$) spin vectors; we have 
chosen $xz$ as the scattering plane with $e^-$ ($e^+$) moving along   
the positive (negative) direction of $z$-axis.  

Insertion of Eqs. (\ref{rhoee})-(\ref{rhoe+}) into Eqs. (\ref{rhopo})
and (\ref{norpo}), neglecting lepton masses, yields
\barr
\rho^{pol}_{\la^{\,}_q,\la^{\,}_{\bar q};\la'_q,\la'_{\bar q}}(\qq)
&=& \frac{1}{4N^{pol}_{\qq}} \, \biggl[ (1 + \cos\alpha_-)\,(1 + \cos\alpha_+)
\> M_{\la^{\,}_{q}, \la^{\,}_{\bar q};+,-}
\> M^*_{\la'_{q}, \la'_{\bar q};+,-}
\nonumber\\
&+& e^{-i(\beta_- +\beta_+)} \, (\sin\alpha_-\sin\alpha_+)
\> M_{\la^{\,}_{q}, \la^{\,}_{\bar q};+,-}
\> M^*_{\la'_{q}, \la'_{\bar q};-,+} \nonumber\\
&+& e^{i(\beta_- +\beta_+)} \, (\sin\alpha_-\sin\alpha_+)
\> M_{\la^{\,}_{q}, \la^{\,}_{\bar q};-,+}
\> M^*_{\la'_{q}, \la'_{\bar q};+,-} \nonumber\\
&+& (1-\cos\alpha_-) \, (1-\cos\alpha_+)
\> M_{\la^{\,}_{q}, \la^{\,}_{\bar q};-,+}
\> M^*_{\la'_{q}, \la'_{\bar q};-,+} \biggl] 
\label{rhopo3}
\earr 
with 
\barr
4N^{pol}_{q\bar q} &=& (1 + \cos\alpha_-) \, (1 + \cos\alpha_+)
\> \left[ \, |M_{+-;+-}|^2 + |M_{-+;+-}|^2 \, \right] \nonumber\\
&+& (1-\cos\alpha_-)\,(1-\cos\alpha_+)\, 
\left[ \, |M_{+-;-+}|^2 + |M_{-+;-+}|^2 \, \right] \nonumber\\
&+& 2 \, \sin\alpha_-\sin\alpha_+ \> \mbox{\rm Re} \left[
e^{-i(\beta_- + \beta_+)} \left( M_{+-;+-} \> M^*_{+-;-+} 
\right. \right.\nonumber\\
&+& \left.\left. M_{-+;+-} \> M^*_{-+;-+} \right) \right]
\label{norpo3} \> .
\earr
In the last equation also quark masses, compared to their energies,
have been neglected.

The explicit expressions of the relevant $e^+e^- \to \qq$ c.m. helicity
amplitudes are given by \cite{abmq}:
\barr
M_{\pm\mp;\pm\mp} = e^2 (1 + \cos\theta) \, [e_q - g_{_Z}(s)
(g_{_V} \mp g_{_A})_l \, (g_{_V} \mp g_{_A})_q] \label{+-+-} \\
M_{\pm\mp;\mp\pm} = e^2 (1 - \cos\theta) \, [e_q - g_{_Z}(s)
(g_{_V} \pm g_{_A})_l \, (g_{_V} \mp g_{_A})_q] \label{+--+} 
\earr
with the usual Standard Model coupling constants:
\barr
g_{_V}^l &=& -{1\over 2} + 2\sin^2\theta_{_W} \quad\quad g_{_A}^l = -{1\over 2} 
\nonumber \\
g_{_V}^{u,c,t} &=&  \>\> {1\over 2} - {4\over 3}\sin^2\theta_{_W} \quad\quad
g_{_A}^{u,c,t} = {1\over 2} \label{cc} \\
g_{_V}^{d,s,b} &=& -{1\over 2} + {2\over 3}\sin^2\theta_{_W} \quad\quad
g_{_A}^{d,s,b} = -{1\over 2}  \nonumber \\
g_{_Z}(s) &=& \frac{1}{4 \sin^2 \theta_{_W} \cos^2 \theta_{_W}}\  
              \frac{s}{(s-M^2_{_Z}) + i M_{_Z} \Gamma_{_Z}} \,\cdot
\nonumber
\earr 

By inserting Eqs. (\ref{+-+-}) and (\ref{+--+}) into Eqs. (\ref{rhopo3})
and (\ref{norpo3}) one obtains: 
\beq
\rho^{pol}_{+-;+-}(q\bar q) = \frac{1}{4N^{pol}_{q\bar q}}
\left[ (1+\cos^2\theta) \,\, F^{pol}_{1,q} + \cos\theta\,\, 
F^{pol}_{2,q} + \sin^2\theta\,\, F^{pol}_{3,q}
\right] \>, \label{eq15}
\eeq
\barr
\rho^{pol}_{+-;-+}(q\bar q) &=& \frac{1}{4N^{pol}_{q\bar q}}
\biggl[ (1+\cos^2\theta)\,(F^{pol}_{4,q} + iF^{pol}_{5,q}) + \cos\theta\, 
(F^{pol}_{6,q} + iF^{pol}_{7,q}) \nonumber\\
&+& \sin^2\theta\, (F^{pol}_{8,q} + iF^{pol}_{9,q}) \biggl] \label{eq16}
\earr
with
\beq
N^{pol}_{q\bar q} = (1+\cos^2\theta) \, F^{pol}_{10,q}
+ \cos\theta \, F^{pol}_{11,q} + \sin^2\theta \, F^{pol}_{12,q} \>.
\label{eq17}
\eeq
The twelve functions $F^{pol}_{i,q}$ depend on the spin directions of the 
incoming leptons: 
\barr
F^{pol}_{1,q} &\equiv&
(1+\cos\alpha_+) \, (1+\cos\alpha_-) \Big[
e_q^2 + |g_{_Z}|^2\,(g_{_V} - g_{_A})_l^2 \, (g_{_V} - g_{_A})_q^2 \nonumber\\
&-& e_q\, 2 \, (\mbox{\rm Re} \, g_{_Z}) \, 
(g_{_V} - g_{_A})_l \, (g_{_V} - g_{_A})_q \Big] \nonumber\\
&+& (1-\cos\alpha_+) \, (1-\cos\alpha_-) \Big[
e_q^2 + |g_{_Z}|^2\,(g_{_V} + g_{_A})_l^2 \, (g_{_V} - g_{_A})_q^2 \nonumber\\
&-& e_q\, 2 \, (\mbox{\rm Re} \, g_{_Z}) \,
(g_{_V} + g_{_A})_l \, (g_{_V} - g_{_A})_q \Big] \nonumber\\
F^{pol}_{2,q} &\equiv&
(1+\cos\alpha_+)\, (1+\cos\alpha_-) \, 2 \, \Big[
e_q^2 + |g_{_Z}|^2\,(g_{_V} - g_{_A})_l^2 \, (g_{_V} - g_{_A})_q^2 \nonumber\\
&-& e_q\, 2 \, (\mbox{\rm Re} \, g_{_Z}) \,
(g_{_V} - g_{_A})_l \, (g_{_V} - g_{_A})_q \Big] \nonumber\\
&-& (1-\cos\alpha_+)\, (1-\cos\alpha_-) \, 2 \, \Big[
e_q^2 + |g_{_Z}|^2\,(g_{_V} + g_{_A})_l^2 \, (g_{_V} - g_{_A})_q^2 \nonumber\\
&-& e_q\, 2\, (\mbox{\rm Re} \, g_{_Z}) \, 
(g_{_V} + g_{_A})_l \, (g_{_V} - g_{_A})_q \Big] \nonumber\\
F^{pol}_{3,q} &\equiv&
2\,\sin\alpha_+ \, \sin\alpha_- \, \Big[
\cos(\beta_+ + \beta_-) \, \Big( e_q^2 + |g_{_Z}|^2 \, 
(g_{_V}^2 - g_{_A}^2)_l \, (g_{_V} - g_{_A})_q^2 \nonumber\\
&-& e_q\,2\,(\mbox{\rm Re}\,g_{_Z}) \, g^l_{_V} \, (g_{_V} - g_{_A})_q \Big)
+ \sin(\beta_+ + \beta_-) \, e_q \, 2 \, (\mbox{\rm Im}\,g_{_Z})\, 
g^l_{_A} \, (g_{_V} - g_{_A})_q \Big] \nonumber\\
F^{pol}_{4,q} &\equiv &
2 \, \sin\alpha_+ \, \sin\alpha_- \Big[
\cos(\beta_+ + \beta_-) [ e_q^2 + |g_{_Z}|^2\, 
(g_{_V}^2 - g_{_A}^2)_l \, (g_{_V}^2 - g_{_A}^2)_q \nonumber\\
&-& e_q\,2\,(\mbox{\rm Re}\,g_{_Z})\, g^l_{_V}\, g^q_{_V} ] +
\sin(\beta_+ + \beta_-) \,e_q\, 2\,
(\mbox{\rm Im}\,g_{_Z})\, g^l_{_A} \, g^q_{_V} \Big] \nonumber\\
F^{pol}_{5,q} &\equiv &
2\,\sin\alpha_+ \,\sin\alpha_- \Big[ \cos(\beta_+ + \beta_-)\,e_q\, 
2(\mbox{\rm Im}\,g_{_Z})\, g^l_{_V} \, g^q_{_A} \nonumber\\
&+& \sin(\beta_+ + \beta_-)
\,e_q\, 2\,(\mbox{\rm Re}\,g_{_Z}) \, g^l_{_A} \, g^q_{_A} \Big] \nonumber\\
F^{pol}_{6,q} &\equiv &
4\,\sin\alpha_+\,\sin\alpha_- \Big[ - \cos(\beta_+ + \beta_-)\,
e_q \,2\,(\mbox{\rm Re}\,g_{_Z})\, g^l_{_A}\,g^q_{_A} \nonumber \\ 
&+& \sin(\beta_+ + \beta_-)\, e_q \, 2\,(\mbox{\rm Im}\,g_{_Z})\, 
g^l_{_V} \, g^q_{_A} \Big] \nonumber\\
F^{pol}_{7,q} &\equiv &
4\,\sin\alpha_+\,\,\sin\alpha_- \Big[ \cos(\beta_+ + \beta_-)\,
e_q \, 2\,(\mbox{\rm Im}\,g_{_Z})\, g^l_{_A}\,g^q_{_V}
- \sin(\beta_+ + \beta_-) \nonumber\\
&\times& [e_q^2 + |g_{_Z}|^2 \, (g_{_V}^2 - g_{_A}^2)_l
\, (g_{_V}^2 - g_{_A}^2)_q - 
e_q \, 2\,(\mbox{\rm Re}\,g_{_Z})\, g_{_V}^l\,\, g_{_V}^q ] \Big] \nonumber\\
F^{pol}_{8,q} &\equiv&
(1+\cos\alpha_+) \, (1+\cos\alpha_-) \Big[
e_q^2 + |g_{_Z}|^2\,(g_{_V} - g_{_A})_l^2 \, (g_{_V}^2 - g_{_A}^2)_q
\nonumber\\ 
&-& e_q\, 2\, (\mbox{\rm Re}\,g_{_Z})\,(g_{_V} - g_{_A})_l\, g^q_{_V}
\Big] \nonumber\\
&+& (1-\cos\alpha_+)\, (1-\cos\alpha_-)
\Big[ e_q^2 + |g_{_Z}|^2\,(g_{_V} + g_{_A})_l^2\,(g_{_V}^2 - g_{_A}^2)_q
\nonumber\\ 
&-& e_q\, 2\, (\mbox{\rm Re}\,g_{_Z})\,(g_{_V} + g_{_A})_l \,g^q_{_V}
\Big] \nonumber\\
F^{pol}_{9,q} &\equiv&
(1+\cos\alpha_+)\, (1+\cos\alpha_-) \Big[
e_q\, 2\,(\mbox{\rm Im}\,g_{_Z})\, (g_{_V} - g_{_A})_l \, g^q_{_A}
\Big] \nonumber\\
&+& (1-\cos\alpha_+)\, (1-\cos\alpha_-) \Big[
e_q\, 2\,(\mbox{\rm Im}\,g_{_Z})\, (g_{_V} + g_{_A})_l \, g^q_{_A}
\Big] \nonumber\\
F^{pol}_{10,q} &\equiv&
(1+\cos\alpha_+)\, (1+\cos\alpha_-) (1/2)\, \Big[
e_q^2 + |g_{_Z}|^2\,(g_{_V} - g_{_A})_l^2\,(g_{_V}^2 + g_{_A}^2)_q 
\nonumber\\
&-& e_q\, 2\, (\mbox{\rm Re}\,g_{_Z})\, (g_{_V} - g_{_A})_l \, g^q_{_V} \Big] 
+ (1-\cos\alpha_+)\, (1-\cos\alpha_-)(1/2) \nonumber\\ 
&\times& \Big[ e_q^2 
+ |g_{_Z}|^2\,(g_{_V} + g_{_A})_l^2\,(g_{_V}^2 + g_{_A}^2)_q 
- e_q\, 2\, (\mbox{\rm Re}\,g_{_Z})\,(g_{_V} + g_{_A})_l \,g^q_{_V}
\Big] \nonumber\\
F^{pol}_{11,q} &\equiv&
(1+\cos\alpha_+)\, (1+\cos\alpha_-) \,2 \,
\Big[ e_q \, (\mbox{\rm Re}\,g_{_Z}) \,(g_{_V} - g_{_A})_l \, g^q_{_A}
\nonumber\\
&-& |g_{_Z}|^2\,(g_{_V} - g_{_A})_l^2\,(g_{_V} g_{_A})_q \Big] \nonumber\\
&-& (1-\cos\alpha_+)\, (1-\cos\alpha_-)\,2 \, 
\Big[ e_q \, (\mbox{\rm Re}\,g_{_Z}) \, (g_{_V} + g_{_A})_l \, g^q_{_A}
\nonumber\\
&-& |g_{_Z}|^2\,(g_{_V} + g_{_A})_l^2\,(g_{_V} g_{_A})_q \Big] \nonumber\\
F^{pol}_{12,q} &\equiv&
(\sin\alpha_+\,\sin\alpha_-) \, \Big[ \cos(\beta_+ + \beta_-)
\,[ e_q^2 - e_q\, 2\,(\mbox{\rm Re}\,g_{_Z})\, g^l_{_V} \, g^q_{_V} 
\nonumber\\
&+& |g_{_Z}|^2\, (g_{_V}^2 - g_{_A}^2)_l \, (g_{_V}^2 + g_{_A}^2)_q] 
+ \sin(\beta_+ + \beta_-) \,[
e_q\, 2\,(\mbox{\rm Im}\,g_{_Z})\, g^l_{_A} \, g^q_{_V}]\,
\Big].
\label{eq18}
\earr

Eqs. (\ref{eq16})-(\ref{eq18}) give the most general expression 
of $\rho^{pol}_{+-;-+}(\qq)$ for a $\qq$ pair obtained in the 
annihilation process of polarized leptons, $e^-e^+ \to \qq$, 
at lowest perturbative order 
in the Standard Model, taking into account both weak and 
electromagnetic interactions ($\gamma$ and $Z_0$ exchanges).

\vskip 12pt
\noindent
{\bf 3 - Numerical values of} {\mbox{\boldmath $\rho_{+-;-+}(\qq)$}}
\vskip 12pt
Let us now consider different polarization states of $e^-$ and $e^+$.
We choose as possible spin directions the 3 coordinate axes, $\hat x, \>
\hat y, \> \hat z$, with spin component $\pm 1/2$ along these directions:
the corresponding values of $(\alpha, \beta$) in Eqs. (\ref{rhoe-})
and (\ref{rhoe+}) are as follows:
\barr 
+\hat x &=& (\pi/2, \> 0) \,\qquad +\hat y = (\pi/2, \> \pi/2) 
\>\>\qquad +\hat z = (0, \> 0) \nonumber\\
-\hat x &=& (\pi/2, \> \pi) \qquad -\hat y = (\pi/2, \> 3\pi/2) 
\qquad -\hat z = (\pi, \> \pi) \label{pol}
\earr
 
We have then a total of $6 \times 6 = 36 $ possible initial spin states.
Many of them will lead to the same value of $\rho^{pol}_{+-;-+}(\qq)$
and it is convenient to group them into the following 9 cases (notice
that Case 3 is just listed for completeness, but it gives identically null 
results due to helicity conservation in the $e^-e^+Z_0$ and 
$e^-e^+\gamma$ vertices):   

\vskip12pt
Case 1: 
\vskip6pt
$\{P(e^-, +\hat z)\, ,\,  P(e^+, +\hat z) \}$

\vskip12pt
Case 2:
\vskip6pt
$\{P(e^-, +\hat z)\, ,\,  P(e^+, +\hat x) \}, \>
 \{P(e^-, +\hat z)\, ,\,  P(e^+, -\hat x) \}, \>
 \{P(e^-, +\hat z)\, ,\,  P(e^+, +\hat y) \},$

$\{P(e^-, +\hat z)\, ,\,  P(e^+, -\hat y) \}, \>
 \{P(e^-, +\hat x)\, ,\,  P(e^+, +\hat z) \}, \>
 \{P(e^-, -\hat x)\, ,\,  P(e^+, +\hat z) \},$ 

$\{P(e^-, +\hat y)\, ,\,  P(e^+, +\hat z) \}, \>
 \{P(e^-, -\hat y)\, ,\,  P(e^+, +\hat z) \}$

\vskip12pt
Case 3:
\vskip6pt
$\{P(e^-, +\hat z)\, ,\,  P(e^+, -\hat z) \}, \> 
 \{P(e^-, -\hat z)\, ,\,  P(e^+, +\hat z) \}$

\vskip12pt
Case 4:
\vskip6pt
$\{P(e^-, -\hat z)\, ,\,  P(e^+, -\hat z) \}$

\vskip12pt
Case 5:
\vskip6pt
$\{P(e^-, +\hat x)\, ,\,  P(e^+, +\hat x) \}, \>
 \{P(e^-, -\hat x)\, ,\,  P(e^+, -\hat x) \}, \> 
 \{P(e^-, +\hat y)\, ,\,  P(e^+, -\hat y) \},$ 

$\{P(e^-, -\hat y)\, ,\,  P(e^+, +\hat y) \}$ 

\vskip12pt
Case 6:
\vskip6pt
$\{P(e^-, +\hat x)\, ,\,  P(e^+, -\hat x) \}, \>
 \{P(e^-, -\hat x)\, ,\,  P(e^+, +\hat x) \}, \>
 \{P(e^-, +\hat y)\, ,\,  P(e^+, +\hat y) \},$

$\{P(e^-, -\hat y)\, ,\,  P(e^+, -\hat y) \}$

\vskip12pt
Case 7:
\vskip6pt
$\{P(e^-, +\hat x)\, ,\,  P(e^+, +\hat y) \}, \>
 \{P(e^-, +\hat y)\, ,\,  P(e^+, +\hat x) \}, \>
 \{P(e^-, -\hat x)\, ,\,  P(e^+, -\hat y) \},$
 
$\{P(e^-, -\hat y)\, ,\,  P(e^+, -\hat x) \}$

\vskip12pt
Case 8:
\vskip6pt
$\{P(e^-, +\hat x)\, ,\,  P(e^+, -\hat y) \}, \>
 \{P(e^-, -\hat y)\, ,\,  P(e^+, +\hat x) \}, \>
 \{P(e^-, -\hat x)\, ,\,  P(e^+, +\hat y) \},$
 
$\{P(e^-, +\hat y)\, ,\,  P(e^+, -\hat x) \}$

\vskip12pt
Case 9:
\vskip6pt
$\{P(e^-, -\hat z)\, ,\,  P(e^+, +\hat x) \}, \>
 \{P(e^-, -\hat z)\, ,\,  P(e^+, +\hat y) \}, \>
 \{P(e^-, -\hat z)\, ,\,  P(e^+, -\hat x) \},$
 
$\{P(e^-, -\hat z)\, ,\,  P(e^+, -\hat y) \}, \>
 \{P(e^-, +\hat x)\, ,\,  P(e^+, -\hat z) \}, \>
 \{P(e^-, -\hat x)\, ,\,  P(e^+, -\hat z) \},$

$\{P(e^-, +\hat y)\, ,\,  P(e^+, -\hat z) \}, \>
 \{P(e^-, -\hat y)\, ,\,  P(e^+, -\hat z) \}$.

\vskip 12pt
The corresponding expressions of the functions $F_{i,q}^{pol}$ are given by:

\vskip12pt
Case 1:
\vskip6pt
$F^{pol, C1}_{1,q} = 4 \Big[
e_q^2 + |g_{_Z}|^2 \, (g_{_V} - g_{_A})_l^2 \, (g_{_V} - g_{_A})_q^2
- e_q \, 2  \,(\mbox{\rm Re}\,g_{_Z}) \, (g_{_V} - g_{_A})_l  
\, (g_{_V} - g_{_A})_q \Big]$ 
\vskip3pt
$F^{pol, C1}_{2,q} = 2\,F^{pol, C1}_{1,q}$
\vskip3pt
$F^{pol, C1}_{3,q} = F^{pol, C1}_{4,q} = F^{pol, C1}_{5,q} 
= F^{pol, C1}_{6,q} = F^{pol, C1}_{7,q} = F^{pol, C1}_{12,q} = 0$
\vskip3pt
$F^{pol, C1}_{8,q} = 4 \Big[
e_q^2 + |g_{_Z}|^2 \, (g_{_V} - g_{_A})_l^2 \, (g_{_V}^2 - g_{_A}^2)_q
- e_q \, 2  \,(\mbox{\rm Re}\,g_{_Z}) \, (g_{_V} - g_{_A})_l \, g_{_V}^q
\Big]$
\vskip3pt
$F^{pol, C1}_{9,q} = 
e_q \, 8 \,(\mbox{\rm Im}\,g_{_Z}) \, (g_{_V} - g_{_A})_l \, g_{_A}^q$
\vskip3pt 
$F^{pol, C1}_{10,q} = 2 \, \Big[
e_q^2 + |g_{_Z}|^2 \, (g_{_V} - g_{_A})_l^2 \, (g_{_V}^2 + g_{_A}^2)_q
- e_q \, 2 \, (\mbox{\rm Re}\,g_{_Z}) \, (g_{_V} - g_{_A})_l \, g_{_V}^q
\Big]$
\vskip3pt
$F^{pol, C1}_{11,q} = 8 \Big[
e_q \, (\mbox{\rm Re}\,g_{_Z}) \, (g_{_V} - g_{_A})_l \, g_{_A}^q
- |g_{_Z}|^2 \, (g_{_V} - g_{_A})_l^2 \,  \,(g_{_V} \, g_{_A})_q
\Big]$ \hfill{(21)}

\vskip12pt
\goodbreak
Case 2:
\nobreak
\vskip6pt
$F^{pol, C2}_{i,q} = (1/2) \, F^{pol, C1}_{i,q} \quad (i = 1-12)$
\hfill{(22)}

\vskip12pt
\goodbreak
Case 3:
\nobreak
\vskip6pt
$F^{pol, C3}_{i,q} = 0 \quad (i = 1-12)$ \hfill{(23)}

\vskip12pt
\goodbreak
Case 4:
\nobreak
\vskip6pt
$F^{pol, C4}_{1,q} = 4 \Big[
e_q^2 + |g_{_Z}|^2 \, (g_{_V} + g_{_A})_l^2 \, (g_{_V} - g_{_A})_q^2
- e_q \, 2 \, (\mbox{\rm Re}\,g_{_Z}) \, (g_{_V} + g_{_A})_l 
\, (g_{_V} - g_{_A})_q \Big]$
\vskip3pt
$F^{pol, C4}_{2,q} = -8 \Big[
e_q^2 + |g_{_Z}|^2 \, (g_{_V} + g_{_A})_l^2 \, (g_{_V} - g_{_A})_q^2
- e_q \, 2 \, (\mbox{\rm Re}\,g_{_Z}) \, (g_{_V} + g_{_A})_l 
\, (g_{_V} - g_{_A})_q \Big]$
\vskip3pt
$F^{pol, C4}_{3,q} = F^{pol, C4}_{4,q} = F^{pol, C4}_{5,q} 
= F^{pol, C4}_{6,q} = F^{pol, C4}_{7,q} = F^{pol, C4}_{12,q} = 0$
\vskip3pt
$F^{pol, C4}_{8,q} = 4 \Big[
e_q^2 + |g_{_Z}|^2 \, (g_{_V} + g_{_A})_l^2 \, (g_{_V}^2 - g_{_A}^2)_q
- e_q  \,2 \, (\mbox{\rm Re}\,g_{_Z}) \, (g_{_V} + g_{_A})_l 
\, g_{_V} ^q \Big]$
\vskip3pt
$F^{pol, C4}_{9,q} = 
e_q \, 8 \, (\mbox{\rm Im}\,g_{_Z}) \, (g_{_V} + g_{_A})_l \, g_{_A}^q$
\vskip3pt
$F^{pol, C4}_{10,q} = 2 \Big[
e_q^2 + |g_{_Z}|^2 \, (g_{_V} + g_{_A})_l^2 \, (g_{_V}^2 + g_{_A}^2)_q
- e_q \, 2  \,(\mbox{\rm Re}\,g_{_Z}) \, (g_{_V} + g_{_A})_l 
\, g_{_V}^q \Big]$
\vskip3pt
$F^{pol, C4}_{11,q} = 8 \Big[
- e_q \, (\mbox{\rm Re}\,g_{_Z})  \, (g_{_V} + g_{_A})_l \, g_{_A}^q
+ |g_{_Z}|^2 \, (g_{_V} + g_{_A})_l^2 \, (g_{_V} g_{_A})_q \Big]$ 
\hfill{(24)}

\vskip12pt
\goodbreak
Case 5:
\nobreak
\vskip6pt
$F^{pol, C5}_{1,q} = 2 \Big[
e_q^2 + |g_{_Z}|^2 \, (g_{_V}^2 + g_{_A}^2)_l \, (g_{_V} - g_{_A})_q^2
- e_q \, 2 \, (\mbox{\rm Re}\,g_{_Z}) \, g_{_V}^l \, (g_{_V} - g_{_A})_q
\Big]$
\vskip3pt
$F^{pol, C5}_{2,q} = 8 \Big[
- |g_{_Z}|^2 \, (g_{_V} g_{_A})_l \, (g_{_V} - g_{_A})_q^2
+ e_q \, (\mbox{\rm Re}\,g_{_Z}) \, g_{_A}^l \, (g_{_V} - g_{_A})_q
\Big]$
\vskip3pt
$F^{pol, C5}_{3,q} = 2 \Big[
e_q^2 + |g_{_Z}|^2 \, (g_{_V}^2 - g_{_A}^2)_l \, (g_{_V} - g_{_A})_q^2
- e_q \, 2 \, (\mbox{\rm Re}\,g_{_Z}) \, g_{_V}^l  \, (g_{_V} - g_{_A})_q 
\Big]$
\vskip3pt
$F^{pol, C5}_{4,q} = 2 \Big[
e_q^2 + |g_{_Z}|^2 \, (g_{_V}^2 - g_{_A}^2)_l \, (g_{_V}^2 - g_{_A}^2)_q
- e_q \, 2  \,(\mbox{\rm Re}\,g_{_Z}) \, g_{_V}^l \, g_{_V}^q
\Big]$
\vskip3pt
$F^{pol, C5}_{5,q} = e_q \, 4 \, (\mbox{\rm Im}\,g_{_Z}) \, g_{_V}^l \, 
g_{_A}^q$
\vskip3pt
$F^{pol, C5}_{6,q} = - e_q \, 8 \, (\mbox{\rm Re}\,g_{_Z}) \, g_{_A}^l \, 
g_{_A}^q$
\vskip3pt
$F^{pol, C5}_{7,q} = e_q \, 8 (\mbox{\rm Im}\,g_{_Z}) \, g_{_A}^l \, g_{_V}^q$
\vskip3pt
$F^{pol, C5}_{8,q} = 2 \Big[
e_q^2 + |g_{_Z}|^2 \, (g_{_V}^2 + g_{_A}^2)_l \, (g_{_V}^2 - g_{_A}^2)_q
- e_q \, 2 \, (\mbox{\rm Re}\,g_{_Z}) \, g_{_V}^l \, g_{_V} ^q \Big]$
\vskip3pt
$F^{pol, C5}_{9,q} = e_q \, 4 \, (\mbox{\rm Im}\,g_{_Z}) \, g_{_V}^l \, 
g_{_A}^q$
\vskip3pt
$F^{pol, C5}_{10,q} =
e_q^2 + |g_{_Z}|^2 \, (g_{_V}^2 + g_{_A}^2)_l \, (g_{_V}^2 + g_{_A}^2)_q
- e_q \, 2 \, (\mbox{\rm Re}\,g_{_Z}) \, g_{_V}^l \, g_{_V}^q$
\vskip3pt
$F^{pol, C5}_{11,q} = 4 \Big[
- e_q (\mbox{\rm Re}\,g_{_Z}) \, g_{_A}^l  g_{_A} ^q
+ |g_{_Z}|^2 \, 2(g_{_V} g_{_A})_l \, (g_{_V} g_{_A})_q \Big]$
\vskip3pt
$F^{pol, C5}_{12,q} = 
e_q^2 + |g_{_Z}|^2 \, (g_{_V}^2 - g_{_A}^2)_l \, (g_{_V}^2 + g_{_A}^2)_q
- e_q \, 2 \, (\mbox{\rm Re}\,g_{_Z}) \, g_{_V}^l \, g_{_V}^q$
\hfill{(25)}

\vskip12pt
\goodbreak
Case 6:
\nobreak
\vskip6pt
$F^{pol, C6}_{1,q} = 2 \Big[
e_q^2 + |g_{_Z}|^2 \, (g_{_V}^2 + g_{_A}^2)_l \, (g_{_V} - g_{_A})_q^2
- e_q \, 2 \, (\mbox{\rm Re}\,g_{_Z}) \, g_{_V}^l \, (g_{_V} - g_{_A})_q
\Big]$
\vskip3pt
$F^{pol, C6}_{2,q} = 8 \Big[
- |g_{_Z}|^2 \, (g_{_V} g_{_A})_l \, (g_{_V} - g_{_A})_q^2
+ e_q \, (\mbox{\rm Re}\,g_{_Z}) \, g_{_A}^l \, (g_{_V} - g_{_A})_q
\Big]$
\vskip3pt
$F^{pol, C6}_{3,q} = -2 \Big[
e_q^2 + |g_{_Z}|^2 \, (g_{_V}^2 - g_{_A}^2)_l \, (g_{_V} - g_{_A})_q^2
- e_q \, 2 \, (\mbox{\rm Re}\,g_{_Z}) \, g_{_V}^l \, (g_{_V} - g_{_A})_q
\Big]$
\vskip3pt
$F^{pol, C6}_{4,q} = -2 \Big[
e_q^2 + |g_{_Z}|^2 \, (g_{_V}^2 - g_{_A}^2)_l \, (g_{_V}^2 - g_{_A}^2)_q
- e_q \, 2 \, (\mbox{\rm Re}\,g_{_Z}) \, g_{_V}^l \, g_{_V}^q \Big]$
\vskip3pt
$F^{pol, C6}_{5,q} =
-e_q \, 4  \,(\mbox{\rm Im}\,g_{_Z}) \, g_{_V}^l \, g_{_A}^q$
\vskip3pt
$F^{pol, C6}_{6,q} =
e_q \, 8 \,(\mbox{\rm Re}\,g_{_Z}) \, g_{_A}^l \, g_{_A}^q$
\vskip3pt
$F^{pol, C6}_{7,q} = 
-e_q \, 8 \, (\mbox{\rm Im}\,g_{_Z}) \, g_{_A}^l \, g_{_V}^q$
\vskip3pt
$F^{pol, C6}_{8,q} = 2 \Big[
e_q^2 + |g_{_Z}|^2 \, (g_{_V}^2 + g_{_A}^2)_l \, (g_{_V}^2 - g_{_A}^2)_q
- e_q \, 2 \, (\mbox{\rm Re}\,g_{_Z}) \, g_{_V}^l \, g_{_V}^q
\Big]$
\vskip3pt
$F^{pol, C6}_{9,q} = 
e_q \, 4 \, (\mbox{\rm Im}\,g_{_Z}) \, g_{_V}^l \, g_{_A}^q$
\vskip3pt
$F^{pol, C6}_{10,q} =
e_q^2 + |g_{_Z}|^2 \, (g_{_V}^2 + g_{_A}^2)_l \, (g_{_V}^2 + g_{_A}^2)_q
- e_q \, 2 \, (\mbox{\rm Re}\,g_{_Z}) \, g_{_V}^l \, g_{_V}^q$
\vskip3pt
$F^{pol, C6}_{11,q} = 4 \Big[
- e_q \, (\mbox{\rm Re}\,g_{_Z}) \, g_{_A}^l \,  g_{_A} ^q
+ |g_{_Z}|^2 \, 2(g_{_V} g_{_A})_l \, (g_{_V} g_{_A})_q \Big]$
\vskip3pt
$F^{pol, C6}_{12,q} =
-e_q^2 - |g_{_Z}|^2 \, (g_{_V}^2 - g_{_A}^2)_l \, (g_{_V}^2 + g_{_A}^2)_q
+ e_q \, 2 \, (\mbox{\rm Re}\,g_{_Z}) \, g_{_V}^l \, g_{_V}^q$
\hfill{(26)}

\vskip12pt
\goodbreak
Case 7:
\nobreak
\vskip6pt
$F^{pol, C7}_{1,q} = 2 \Big[
e_q^2 + |g_{_Z}|^2 \, (g_{_V}^2 + g_{_A}^2)_l \, (g_{_V} - g_{_A})_q^2
- e_q \, 2 \, (\mbox{\rm Re}\,g_{_Z}) \, g_{_V}^l \,(g_{_V} - g_{_A})_q
\Big]$
\vskip3pt
$F^{pol, C7}_{2,q} = 4 \Big[
- |g_{_Z}|^2 \, 2 (g_{_V} g_{_A})_l \,(g_{_V} - g_{_A})_q^2
+ e_q \, 2 \, (\mbox{\rm Re}\,g_{_Z}) \, g_{_A}^l \, (g_{_V} - g_{_A})_q
\Big]$
\vskip3pt
$F^{pol, C7}_{3,q} =
e_q \, 4 \, (\mbox{\rm Im}\,g_{_Z}) \, g_{_A}^l \, (g_{_V} - g_{_A})_q$
\vskip3pt
$F^{pol, C7}_{4,q} =
e_q \, 4 \, (\mbox{\rm Im}\,g_{_Z}) \, g_{_A}^l \, g_{_V}^q$
\vskip3pt
$F^{pol, C7}_{5,q} =
e_q \, 4 \, (\mbox{\rm Re}\,g_{_Z}) \, g_{_A}^l \, g_{_A}^q$
\vskip3pt
$F^{pol, C7}_{6,q} =
e_q  \,8 \, (\mbox{\rm Im}\,g_{_Z}) \, g_{_V}^l \, g_{_A}^q$
\vskip3pt
$F^{pol, C7}_{7,q} = -4 \Big[
e_q^2 + |g_{_Z}|^2 \, (g_{_V}^2 - g_{_A}^2)_l \, (g_{_V}^2 - g_{_A}^2)_q
- e_q \, 2 \, (\mbox{\rm Re}\,g_{_Z}) \, g_{_V}^l \, g_{_V}^q \Big]$
\vskip3pt
$F^{pol, C7}_{8,q} = 2 \Big[
e_q^2 + |g_{_Z}|^2 \, (g_{_V}^2 + g_{_A}^2)_l \, (g_{_V}^2 - g_{_A}^2)_q
- e_q \, 2 \, (\mbox{\rm Re}\,g_{_Z}) \, g_{_V}^l \, g_{_V}^q \Big]$
\vskip3pt
$F^{pol, C7}_{9,q} = 
e_q \, 4 \, (\mbox{\rm Im}\,g_{_Z}) \, g_{_V}^l \, g_{_A}^q$
\vskip3pt
$F^{pol, C7}_{10,q} =
e_q^2 + |g_{_Z}|^2 \, (g_{_V}^2 + g_{_A}^2)_l \,(g_{_V}^2 + g_{_A}^2)_q
- e_q \, 2 \, (\mbox{\rm Re}\,g_{_Z}) \, g_{_V}^l \, g_{_V}^q$
\vskip3pt
$F^{pol, C7}_{11,q} = 4 \Big[
- e_q \, (\mbox{\rm Re}\,g_{_Z}) \, g_{_A}^l \,  g_{_A}^q
+ |g_{_Z}|^2 \, 2\,(g_{_V} g_{_A})_l \, (g_{_V} g_{_A})_q \Big]$
\vskip3pt
$F^{pol, C7}_{12,q} =
e_q \, 2 \, (\mbox{\rm Im}\,g_{_Z}) \, g_{_A}^l \, g_{_V}^q$
\hfill{(27)}

\vskip12pt
\goodbreak
Case 8:
\nobreak
\vskip6pt
$F^{pol, C8}_{1,q} = 2 \Big[
e_q^2 + |g_{_Z}|^2 \, (g_{_V}^2 + g_{_A}^2)_l \, (g_{_V} - g_{_A})_q^2
- e_q \, 2 \, (\mbox{\rm Re}\,g_{_Z}) \, g_{_V}^l \, (g_{_V} - g_{_A})_q
\Big]$
\vskip3pt
$F^{pol, C8}_{2,q} = 8 \Big[
- |g_{_Z}|^2 \, (g_{_V} g_{_A})_l \, (g_{_V} - g_{_A})_q^2
+ e_q \, (\mbox{\rm Re}\,g_{_Z}) \,  g_{_A}^l \, (g_{_V} - g_{_A})_q
\Big]$
\vskip3pt
$F^{pol, C8}_{3,q} =
-e_q \, 4 \, (\mbox{\rm Im}\,g_{_Z}) \,g _{_A}^l \, (g_{_V} - g_{_A})_q$
\vskip3pt
$F^{pol, C8}_{4,q} =
-e_q \, 4 \, (\mbox{\rm Im}\,g_{_Z}) \, g_{_A}^l \, g_{_V}^q$
\vskip3pt
$F^{pol, C8}_{5,q} =
-e_q \, 4 \, (\mbox{\rm Re}\,g_{_Z}) \, g_{_A}^l \, g_{_A}^q$
\vskip3pt
$F^{pol, C8}_{6,q} =
-e_q \, 8 \, (\mbox{\rm Im}\,g_{_Z}) \, g_{_V}^l \, g_{_A}^q$ 
\vskip3pt
$F^{pol, C8}_{7,q} = 4 \Big[
e_q^2 + |g_{_Z}|^2 \, (g_{_V}^2 - g_{_A}^2)_l \, (g_{_V}^2 - g_{_A}^2)_q
- e_q \, 2 \, (\mbox{\rm Re}\,g_{_Z}) \, g_{_V}^l \, g_{_V}^q
\Big]$
\vskip3pt
$F^{pol, C8}_{8,q} = 2 \Big[
e_q^2 + |g_{_Z}|^2 \, (g_{_V}^2 + g_{_A}^2)_l \, (g_{_V}^2 - g_{_A}^2)_q
- e_q \, 2 \, (\mbox{\rm Re}\,g_{_Z}) \, g_{_V}^l \, g_{_V}^q
\Big]$
\vskip3pt
$F^{pol, C8}_{9,q} =
e_q \, 4 \, (\mbox{\rm Im}\,g_{_Z}) \,g_{_V}^l \, g_{_A}^q$
\vskip3pt
$F^{pol, C8}_{10,q} =
e_q^2 + |g_{_Z}|^2 \, (g_{_V}^2 + g_{_A}^2)_l \, (g_{_V}^2 + g_{_A}^2)_q
- e_q \, 2 \, (\mbox{\rm Re}\,g_{_Z}) \, g_{_V}^l \, g_{_V}^q$
\vskip3pt
$F^{pol, C8}_{11,q} = 4 \Big[
- e_q \, (\mbox{\rm Re}\,g_{_Z}) \, g_{_A}^l \,  g_{_A}^q
+ |g_{_Z}|^2 \, 2(g_{_V} g_{_A})_l \, (g_{_V} g_{_A})_q \Big]$
\vskip3pt
$F^{pol, C8}_{12,q} =
-e_q \, 2 \, (\mbox{\rm Im}\,g_{_Z}) \, g_{_A}^l \, g_{_V}^q$
\hfill{(28)}

\vskip12pt
\goodbreak
Case 9:
\nobreak
\vskip6pt
$F^{pol, C9}_{i,q} = (1/2) \, F^{pol, C4}_{i,q} \quad (i = 1-12)$
\hfill{(29)}

\setcounter{equation}{29}
\vskip12pt
We can now compute $\rho^{pol}_{+-;-+}(\qq)$ for any initial lepton spin
state, and at any energy, 
by using Eqs. (21)-(29), together with Eq. (\ref{cc}), in Eqs. (17)
and (18). We do it here first at the $Z_0$ pole, $\sqrt s = M_{_Z}$, where
\beq
g_{_Z}(s=M_{_Z}^2) =
-i \,\frac{M_{_Z}/\Gamma_{_Z}}{4\,\sin^2\theta_{_W}\,
 \cos^2\theta_{_W}} \> \cdot
\label{eq30}
\eeq
Taking \cite{pdg} $\sin^2\theta_{_W} = 0.231$, $M_{_Z} = 91.187$ GeV/$c^2$ and
$\Gamma_{_Z} = 2.490$ GeV yields for $u$-type quarks:
\barr
\rho_{+-;-+}^{pol, C1,C2}(u\bar u; \sqrt s = M_{_Z}) &=&
-0.369\,(1+i\, 0.132)\,%%\bigg[
\frac{\sin^2\theta}{1+ \cos^2\theta - 1.335\,\cos\theta} %%\bigg]
\nonumber\\ \nonumber\\
\rho_{+-;-+}^{pol, C4,C9}(u\bar u; \sqrt s = M_{_Z}) &=&
-0.370\,(1-i\, 0.113)\,%%\bigg[
\frac{\sin^2\theta}{1+ \cos^2\theta + 1.336\,\cos\theta} %%\bigg]
\nonumber\\ \nonumber\\
\mbox{\rm Re}\,[\rho_{+-;-+}^{pol, C5}(u\bar u; \sqrt s = M_{_Z})] &=&
-0.371 \, %%\bigg[
\frac{0.003 - \cos^2\theta}
{0.008 + \cos^2\theta + 0.102\,\cos\theta} %%\bigg]
\nonumber\\ \nonumber\\
\mbox{\rm Im}\,[\rho_{+-;-+}^{pol, C5}(u\bar u; \sqrt s = M_{_Z})] &=&
+0.371 \, %%\bigg[
\frac{0.009 + 0.047\,\cos\theta}
{0.008 + \cos^2\theta + 0.102\,\cos\theta} %%\bigg]
\nonumber\\ \nonumber\\
\mbox{\rm Re}\,[\rho_{+-;-+}^{pol, C6}(u\bar u; \sqrt s = M_{_Z})] &=&
-0.371 \, %%\bigg[
\frac{1 - 0.003\,\cos^2\theta}
{1 + 0.008\,\cos^2\theta + 0.102\,\cos\theta} %%\bigg]
\nonumber\\ \nonumber\\
\mbox{\rm Im}\,[\rho_{+-;-+}^{pol, C6}(u\bar u; \sqrt s = M_{_Z})] &=&
-0.371 \, %\bigg[
\frac{0.009\,\cos^2\theta + 0.047\,\cos\theta}
{1 + 0.008\,\cos^2\theta + 0.102\,\cos\theta} %\bigg]
\nonumber\\ \nonumber\\
\mbox{\rm Re}\,[\rho_{+-;-+}^{pol, C7}(u\bar u; \sqrt s = M_{_Z})] &=&
-0.374 \, %\bigg[
\frac{0.911 - \cos^2\theta - 0.018\,\cos\theta}
{1+ 0.934\,\cos^2\theta + 0.195\,\cos\theta} %\bigg]
\nonumber\\ \nonumber\\
\mbox{\rm Im}\,[\rho_{+-;-+}^{pol, C7}(u\bar u; \sqrt s = M_{_Z})] &=&
+0.374 \, %\bigg[
\frac{0.009\,\sin^2\theta - 1.901\,\cos\theta}
{1+ 0.934\,\cos^2\theta + 0.195\,\cos\theta} %\bigg]
\nonumber\\ \nonumber\\
\mbox{\rm Re}\,[\rho_{+-;-+}^{pol, C8}(u\bar u; \sqrt s = M_{_Z})] &=&
-0.374 \, %\bigg[
\frac{1 -0.911\, \cos^2\theta + 0.018\,\cos\theta}
{0.934 + \cos^2\theta + 0.195\,\cos\theta} %\bigg]
\nonumber\\ \nonumber\\
\mbox{\rm Im}\,[\rho_{+-;-+}^{pol, C8}(u\bar u; \sqrt s = M_{_Z})] &=&
+0.374 \, %\bigg[
\frac{0.009\,\sin^2\theta + 1.901\,\cos\theta}
{0.934 + \cos^2\theta + 0.195\,\cos\theta} %\bigg]
\label{eq31}
\earr
and for $d$-type quarks
\barr
\rho_{+-;-+}^{pol, C1,C2}(d\bar d; \sqrt s = M_{_Z}) &=&
-0.176\,(1 + i\, 0.108)\, %\bigg[
\frac{\sin^2\theta}{1+ \cos^2\theta - 1.871\,\cos\theta} %\bigg]
\nonumber\\ \nonumber\\
\rho_{+-;-+}^{pol, C4,C9}(d\bar d; \sqrt s = M_{_Z}) &=&
-0.176\,(1-i\, 0.092)\,%\bigg[
\frac{\sin^2\theta}{1+ \cos^2\theta + 1.871\,\cos\theta} %\bigg]
\nonumber\\ \nonumber\\
\mbox{\rm Re}\,[\rho_{+-;-+}^{pol, C5}(d\bar d; \sqrt s = M_{_Z})] &=&
-0.176 \, %\bigg[
\frac{0.004 - \cos^2\theta}
{0.006 + \cos^2\theta + 0.142\,\cos\theta} %\bigg]
\nonumber\\ \nonumber\\
\mbox{\rm Im}\,[\rho_{+-;-+}^{pol, C5}(d\bar d; \sqrt s = M_{_Z})] &=&
+0.176 \, %\bigg[
\frac{0.008 + 0.069\,\cos\theta}
{0.006 + \cos^2\theta + 0.142\,\cos\theta} %\bigg]
\nonumber\\ \nonumber\\
\mbox{\rm Re}\,[\rho_{+-;-+}^{pol, C6}(d\bar d; \sqrt s = M_{_Z})] &=&
-0.176 \, %\bigg[
\frac{1 - 0.004\,\cos^2\theta}
{1+ 0.006\,\cos^2\theta + 0.142\,\cos\theta} %\bigg]
\nonumber\\ \nonumber\\
\mbox{\rm Im}\,[\rho_{+-;-+}^{pol, C6}(d\bar d; \sqrt s = M_{_Z})] &=&
-0.176 \, %\bigg[
\frac{0.008\,\cos^2\theta + 0.069\,\cos\theta}
{1+ 0.006\,\cos^2\theta + 0.142\,\cos\theta} %\bigg]
\nonumber\\ \nonumber\\
\mbox{\rm Re}\,[\rho_{+-;-+}^{pol, C7}(d\bar d; \sqrt s = M_{_Z})] &=&
-0.184 \, %\bigg[
\frac{0.872 - \cos^2\theta - 0.014\,\cos\theta}
{1+ 0.953\,\cos^2\theta + 0.276\,\cos\theta} %\bigg]
\nonumber\\ \nonumber\\
\mbox{\rm Im}\,[\rho_{+-;-+}^{pol, C7}(d\bar d; \sqrt s = M_{_Z})] &=&
+0.184 \, %\bigg[
\frac{0.007\,\sin^2\theta - 1.855\,\cos\theta}
{1+ 0.953\,\cos^2\theta + 0.276\,\cos\theta} %\bigg]
\nonumber\\ \nonumber\\
\mbox{\rm Re}\,[\rho_{+-;-+}^{pol, C8}(d\bar d; \sqrt s = M_{_Z})] &=&
-0.184 \, %\bigg[
\frac{1 - 0.872\, \cos^2\theta + 0.014\,\cos\theta}
{0.953 + \cos^2\theta + 0.276\,\cos\theta} %\bigg]
\nonumber\\ \nonumber\\
\mbox{\rm Im}\,[\rho_{+-;-+}^{pol, C8}(d\bar d; \sqrt s = M_{_Z})] &=&
+0.184 \, %\bigg[
\frac{0.007\,\sin^2\theta + 1.855\,\cos\theta}
{0.953 + \cos^2\theta + 0.276\,\cos\theta} %\bigg]\> \cdot
\label{eq32}
\earr

At lower energies, instead, where one can neglect all weak interactions
[that is, setting $g_{_Z} = 0$ in Eqs. (21)-(29) and taking into account
quark masses] one obtains for any flavour:
\barr
\rho_{+-;-+}^{pol, C1,C2,C4,C9}(\qq; \sqrt s \ll M_{_Z}) &=&
{1 \over 2} \, \frac{\sin^2\theta}{1+ \cos^2\theta+\epsilon^2\sin^2\theta}
\nonumber \\
\rho_{+-;-+}^{pol, C5}(\qq; \sqrt s \ll M_{_Z}) &=& {1 \over 2}
\nonumber \\ 
\rho_{+-;-+}^{pol, C6}(\qq; \sqrt s \ll M_{_Z}) &=& - {1 \over 2}
\frac{\cos^2\theta}{\cos^2\theta+\epsilon^2\sin^2\theta}
\nonumber \\ 
\mbox{\rm Re}\,[\rho_{+-;-+}^{pol, C7,C8}(\qq; \sqrt s \ll M_{_Z})] &=&
{1 \over 2} \, \frac{\sin^2\theta}{1+ \cos^2\theta+\epsilon^2\sin^2\theta}
\nonumber \\
\mbox{\rm Im}\,[\rho_{+-;-+}^{pol, C7}(\qq; \sqrt s \ll M_{_Z})] &=&
\frac{-\cos\theta}{1+ \cos^2\theta+\epsilon^2\sin^2\theta}
\nonumber \\
\mbox{\rm Im}\,[\rho_{+-;-+}^{pol, C8}(\qq; \sqrt s \ll M_{_Z})] &=&
\frac{\cos\theta}{1+ \cos^2\theta+\epsilon^2\sin^2\theta} \> ,
\label{eq33} 
\earr
where $\epsilon = 2m_q/\sqrt s$, which, for heavy flavours, might not be
negligible at $\sqrt s \ll M_{_Z}$.  

Insertion of Eqs. (31) and (32) or (33) into Eq. (1) allows to give predictions 
for the relation between $\rho_{1,-1}(V)$ and $\rho_{0,0}(V)$, both of them
measurable quantities. Eq. (1) holds for vector mesons with a large energy 
fraction $x_{_E}$ and collinear with the parent jet; $q$ is the  
quark flavour which contributes dominantly to the final vector meson
production ({\it e.g.} $c$ in $D^*$); an average should be taken if 
more than one flavour contributes \cite{abmq}.

Notice that we expect \cite{abmq} $\rho_{0,0}(V)$ to be independent
of the production angle $\theta$, so that the sign of $\rho_{1,-1}(V)$
and its $\theta$ dependence are entirely given by the elementary dynamics,
via $\rho_{+-;-+}(\qq)$; for unpolarized $e^+$ and $e^-$ such dynamics
is given by Eq. (4) or (5), and for polarized ones by Eqs. (31) and (32) 
or (33). We turn now to a discussion of these equations and a comparison with 
the unpolarized case. 

\vskip 12pt
\noindent
{\bf 4 - Comments and conclusions}
\vskip 12pt
We show our numerical results for $\rho^{pol}_{+-;-+}(\qq)$ in 
Figs. 1-6. We give results only for those cases which strongly
differ from the unpolarized case and have such peculiar features which
would make a measurement of $\rho_{1,-1}(V)$ in agreement with them
an unquestionable test of our approach. In Figs. 1-4 we consider
the LEP high energy case, $\sqrt s = M_{_Z}$, and in Figs. 5-6 the
lower energy case, $\sqrt s \ll M_{_Z}$. 
  
In Fig. 1 we plot as functions of $\theta$ (the $V$ production angle in 
the $e^-e^+$ c.m. frame) the real part of $\rho^{pol}_{+-;-+}(u\bar u)$
at LEP energy for cases: $C5$, $C6$, $C1,2$ and $C4,9$. Also the value of
$\rho_{+-;-+}(u\bar u)$ for unpolarized leptons 
is reported for comparison. In Fig. 2 we do the same for 
$d$-type quarks.

In Fig. 3 we plot the imaginary part of $\rho^{pol}_{+-;-+}(u\bar u)$
at LEP energy for cases: $C5$, $C7$ and $C8$. In all other cases, 
{\it including the unpolarized one}, such imaginary part is much smaller 
and should lead to a measurement of Im $\rho_{1,-1}(V) \simeq 0$.
The same is done in Fig. 4 for $d$, $s$ and $b$ quarks. 

In Fig. 5 we plot the real part of $\rho^{pol}_{+-;-+}(\qq;\sqrt s \ll M_{_Z})$ 
taking into account only electromagnetic interactions 
for cases $C5$ and $C6$. All other cases give the same result 
as unpolarized leptons, result which is reported for comparison. Quark masses 
have been taken into account, setting $\epsilon = 2m_q/\sqrt s = 0.1$.

In Fig. 6 we plot the imaginary part of 
$\rho^{pol}_{+-;-+}(\qq;\sqrt s \ll M_{_Z})$ taking into account 
only electromagnetic interactions (and quark masses, $\epsilon = 0.1$)
for cases $C7$ and $C8$. In all other cases, {\it including the unpolarized 
one}, the imaginary part is zero. 

Figs. 1-6 show beyond any possible doubt how the elementary dynamics
might lead to very different values of $\rho_{1,-1}(V)$, according 
to the different spin states of the initial $e^+$ and $e^-$. 
A measurement in agreement with our predictions would confirm in a 
definite way the necessity of coherent effects in the quark fragmentation 
and prove all subtleties of the Standard Model dynamics.

Let us further comment on the most typical cases. The 
possible spin configurations and the definitions of the various 
cases are listed at the beginning of Section 3. Concerning the real parts 
at LEP energy -- Figs. 1 and 2 -- case $C5$ presents the most striking
features, both in sign and $\theta$ dependence and shows a drastic 
difference from the unpolarized case; also $C6$ has a peculiar, almost 
constant, $\theta$ dependence which should be easily detectable.
These two cases correspond to $e^+$ and $e^-$ transversely polarized
in the same direction, with either parallel or opposite spins.
Cases $C1,2$ and $C4,9$ also deviate largely
from the unpolarized case, in particular for charge -1/3 quarks: $C1$
and $C4$ correspond to initial leptons with opposite helicities
and $C2$, $C9$ to spin configurations in which one of the
lepton is longitudinally polarized and the other is transversely 
polarized. 

Cases $C7$ and $C8$, leptons transversely polarized in 
different directions, lead to results similar to unpolarized leptons
for the real part of $\rho^{pol}_{+-;-+}(\qq)$; however, contrary to the
unpolarized case, they give large values, strongly varying with $\theta$
-- Figs. 3 and 4 -- for Im $\rho^{pol}_{+-;-+}(\qq)$, 
which makes them very interesting. Also $C5$ exhibits a peculiar 
$\theta$ dependence in Im $\rho^{pol}_{+-;-+}(\qq)$

At lower energy, when only electromagnetic interactions contribute,
cases $C5$ and $C6$ are simple and very interesting -- see Fig. 5 -- for the 
real parts of $\rho^{pol}_{+-;-+}(\qq)$; cases $C7$ and $C8$ are unique
providers of sizeable imaginary parts of $\rho^{pol}_{+-;-+}(\qq)$,
Fig. 6. 

We have thus completed the study of the off-diagonal helicity density
matrix element $\rho_{1,-1}(V)$ of vector mesons produced in 
$e^+e^-$ annihilations into two jets, selecting vector 
mesons with a large energy fraction and small transverse momentum
inside one of the jets. The idea was suggested in Refs. \cite{akp}
and \cite{aamr}, and the first numerical predictions, given in 
Ref. \cite{abmq}, have been confirmed by some experimental data 
\cite{opal1,opal2}.
We have considered here the most general case of polarized $e^+$ and
$e^-$; we have given numerical results both at LEP energy, $\sqrt s = M_{_Z}$,
and for $\sqrt s \ll M_{_Z}$,
but our formulae, Eqs. (17)-(19) and (21)-(29), are valid at any energy 
and take into account both electromagnetic and weak interactions.  

At the moment, there is no operating $e^+e^-$ collider with polarized
beams; however, future generations of linear colliders are being
planned and our study might indicate very good reasons to seriously
consider polarization options. 
 
\vskip 24pt
\noindent
{\bf Acknowledgements}
\vskip 6pt
P. Q. is grateful to FAPESP of Brazil for financial support.
M. B. would like to thank the Department of Theoretical Physics 
of Universit\`a di Torino where this work was initiated.
The work of M. B. is supported in part by the EU Fourth Framework Programme
`Training and Mobility of Researchers', Network `Quantum
Chromodynamics and the Deep Structure of Elementary Particles', contract
FMRX-CT98-0194 (DG 12 - MIHT).

%\end{document}
%%%%%%%%%%%%%%%%%%%%%%%%%%%%%%%%%%%%%%%%%%%%%%%%%%%%%%%%%%%%%%%%%%%%%%%%%%%%%%%

\vspace{18pt}

\noindent
{\bf Figure Captions}

\vskip 12pt
\noindent
{\bf Fig. 1} - Plot of Re[$\rho^{pol}_{+-;-+}(u\bar u; \sqrt s = M_{_Z})$]
as a function of $\theta$ (the production angle of the vector meson in 
the $e^-e^+$ c.m. frame) for cases: $C5$, $C6$ (both leptons transversely
polarized with spins either parallel or anti-parallel); $C1$, $C4$ (leptons
with opposite helicities); $C2$, $C9$ (one lepton longitudinally polarized, 
the other transversely polarized). Also the value of 
$\rho_{+-;-+}(u\bar u; \sqrt s = M_{_Z})$ for unpolarized leptons is shown 
for comparison. In all other cases one obtains results similar to
the unpolarized case.

\vskip 12pt
\noindent
{\bf Fig. 2} - The same as in Fig. 1, for $d$-type quarks.

\vskip 12pt
\noindent
{\bf Fig. 3} - Plot of Im[$\rho^{pol}_{+-;-+}(u\bar u; \sqrt s = M_{_Z})$]
as a function of $\theta$ (the production angle of the vector meson in 
the $e^-e^+$ c.m. frame) for cases: $C5$ (both leptons transversely
polarized with spins either parallel or anti-parallel);
$C7$, $C8$ (both leptons transversely polarized, in different directions).
In all other cases, {\it including the unpolarized one}, 
Im[$\rho^{pol}_{+-;-+}(u\bar u; \sqrt s = M_{_Z})]\simeq 0$.

\vskip 12pt
\noindent
{\bf Fig. 4} - The same as in Fig. 3, for $d$-type quarks.

\vskip 12pt
\noindent
{\bf Fig. 5} - Plot of Re[$\rho^{pol}_{+-;-+}(\qq; \sqrt s \ll M_{_Z})$] 
as a function of $\theta$ (the production angle of the vector meson in 
the $e^-e^+$ c.m. frame) for cases $C5$ and $C6$ 
(both leptons transversely polarized with spins either parallel or 
anti-parallel). All other cases give the same result given by 
unpolarized leptons, which is shown for comparison. Quark masses 
have been taken into account, with $\epsilon = 2m_q/\sqrt s = 0.1$.

\vskip 12pt
\noindent
{\bf Fig. 6} - Plot of Im[$\rho^{pol}_{+-;-+}(\qq; \sqrt s \ll M_{_Z})$] 
for cases $C7$ and $C8$ (both leptons transversely polarized, in
different directions). In all other cases, {\it including the unpolarized 
one}, Im[$\rho^{pol}_{+-;-+}(\qq; \sqrt s \ll M_{_Z})] = 0$. 
Again, $\epsilon = 0.1$.

\clearpage

\begin{figure}[c]
\centerline{
\epsfig{figure=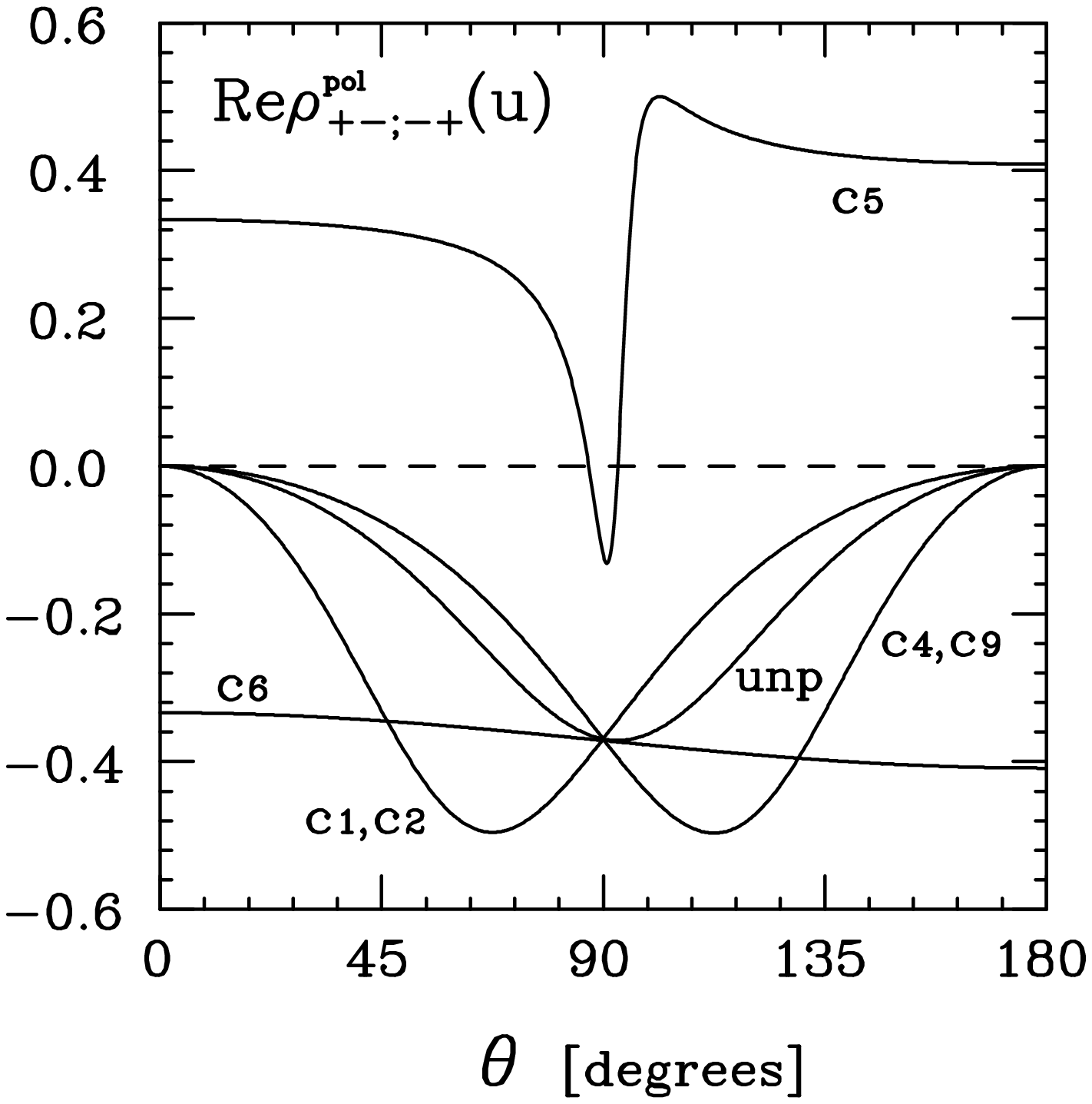,bbllx=50pt,bblly=200pt,bburx=530pt,%
bbury=650pt,width=15.0cm,height=15.0cm}}
 \begin{center}
 \vspace{12pt}
 \begin{minipage}[c]{13cm}
 {\small {\bf Fig. 1:}
Plot of Re[$\rho^{pol}_{+-;-+}(u\bar u; \sqrt s = M_{_Z})$]
as a function of $\theta$ (the production angle of the vector meson in 
the $e^-e^+$ c.m. frame) for cases: $C5$, $C6$ (both leptons transversely
polarized with spins either parallel or anti-parallel); $C1$, $C4$ (leptons
with opposite helicities); $C2$, $C9$ (one lepton longitudinally polarized, 
the other transversely polarized). Also the value of 
$\rho_{+-;-+}(u\bar u; \sqrt s = M_{_Z})$ for unpolarized leptons is shown 
for comparison. In all other cases one obtains results similar to
the unpolarized case. }
 \end{minipage}
 \end{center}
\end{figure}

\clearpage

\begin{figure}[c]
\centerline{
\epsfig{figure=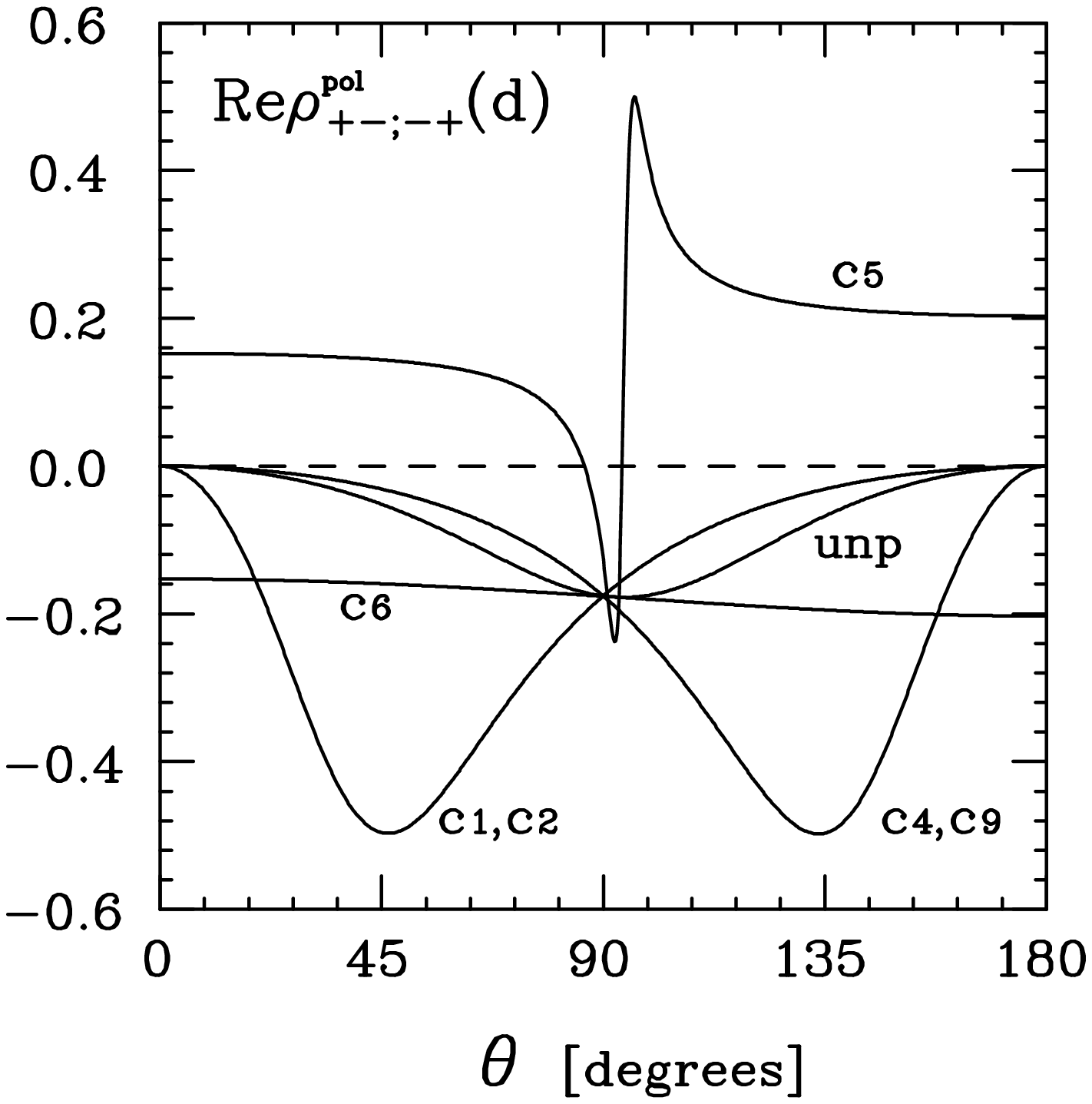,bbllx=50pt,bblly=200pt,bburx=530pt,%
bbury=650pt,width=15.0cm,height=15.0cm}}
 \begin{center}
 \vspace{12pt}
 \begin{minipage}[c]{13cm}
 {\small {\bf Fig. 2:}
The same as in Fig. 1, for $d$-type quarks. }
 \end{minipage}
 \end{center}
\end{figure}

\clearpage

\begin{figure}[c]
\centerline{
\epsfig{figure=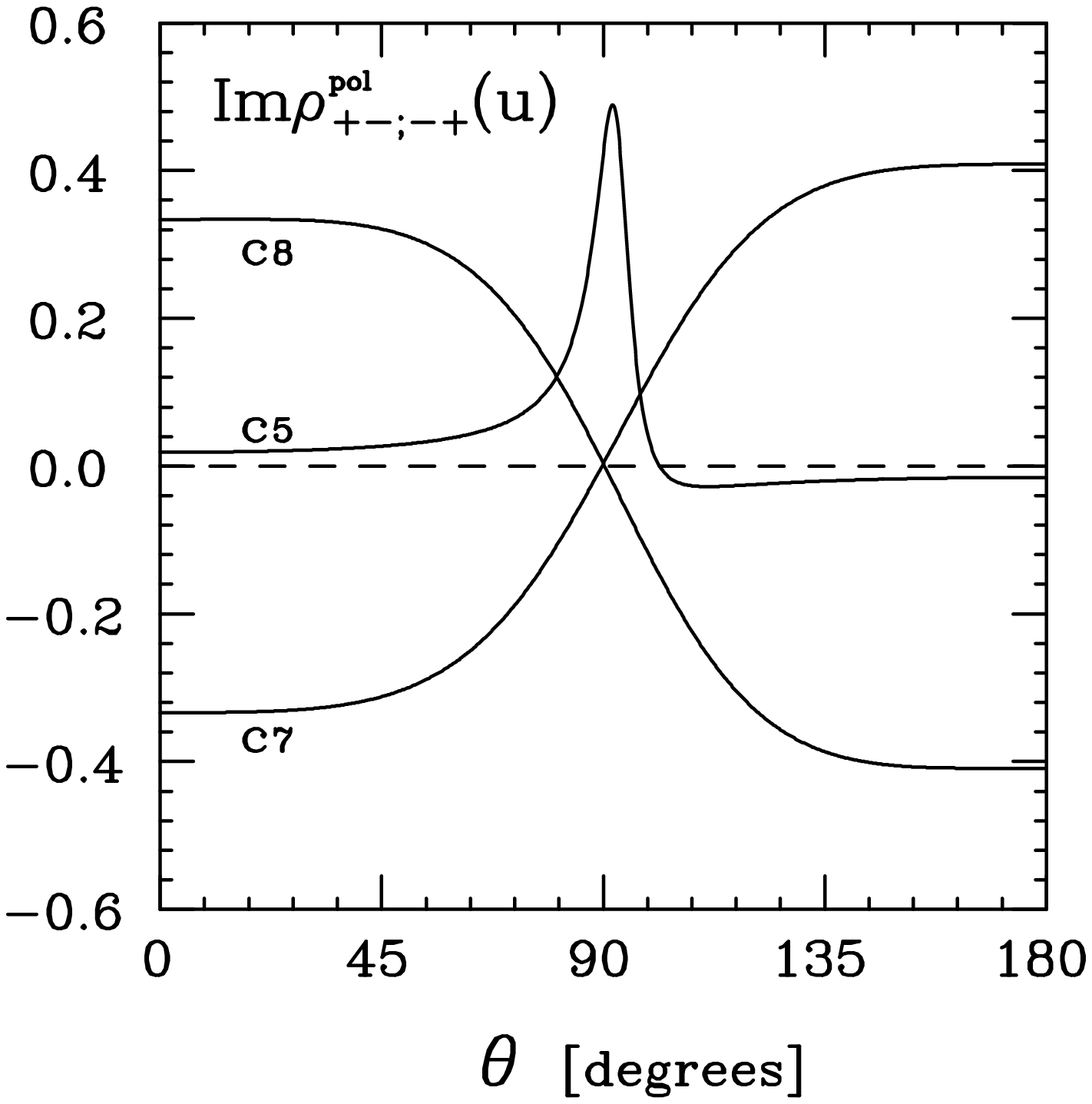,bbllx=50pt,bblly=200pt,bburx=530pt,%
bbury=650pt,width=15.0cm,height=15.0cm}}
 \begin{center}
 \vspace{12pt}
 \begin{minipage}[c]{13cm}
 {\small {\bf Fig. 3:}
Plot of Im[$\rho^{pol}_{+-;-+}(u\bar u; \sqrt s = M_{_Z})$]
as a function of $\theta$ (the production angle of the vector meson in 
the $e^-e^+$ c.m. frame) for cases: $C5$ (both leptons transversely
polarized with spins either parallel or anti-parallel);
$C7$, $C8$ (both leptons transversely polarized, in different directions).
In all other cases, {\it including the unpolarized one}, 
Im[$\rho^{pol}_{+-;-+}(u\bar u; \sqrt s = M_{_Z})]\simeq 0$. }
 \end{minipage}
 \end{center}
\end{figure}

\clearpage

\begin{figure}[c]
\centerline{
\epsfig{figure=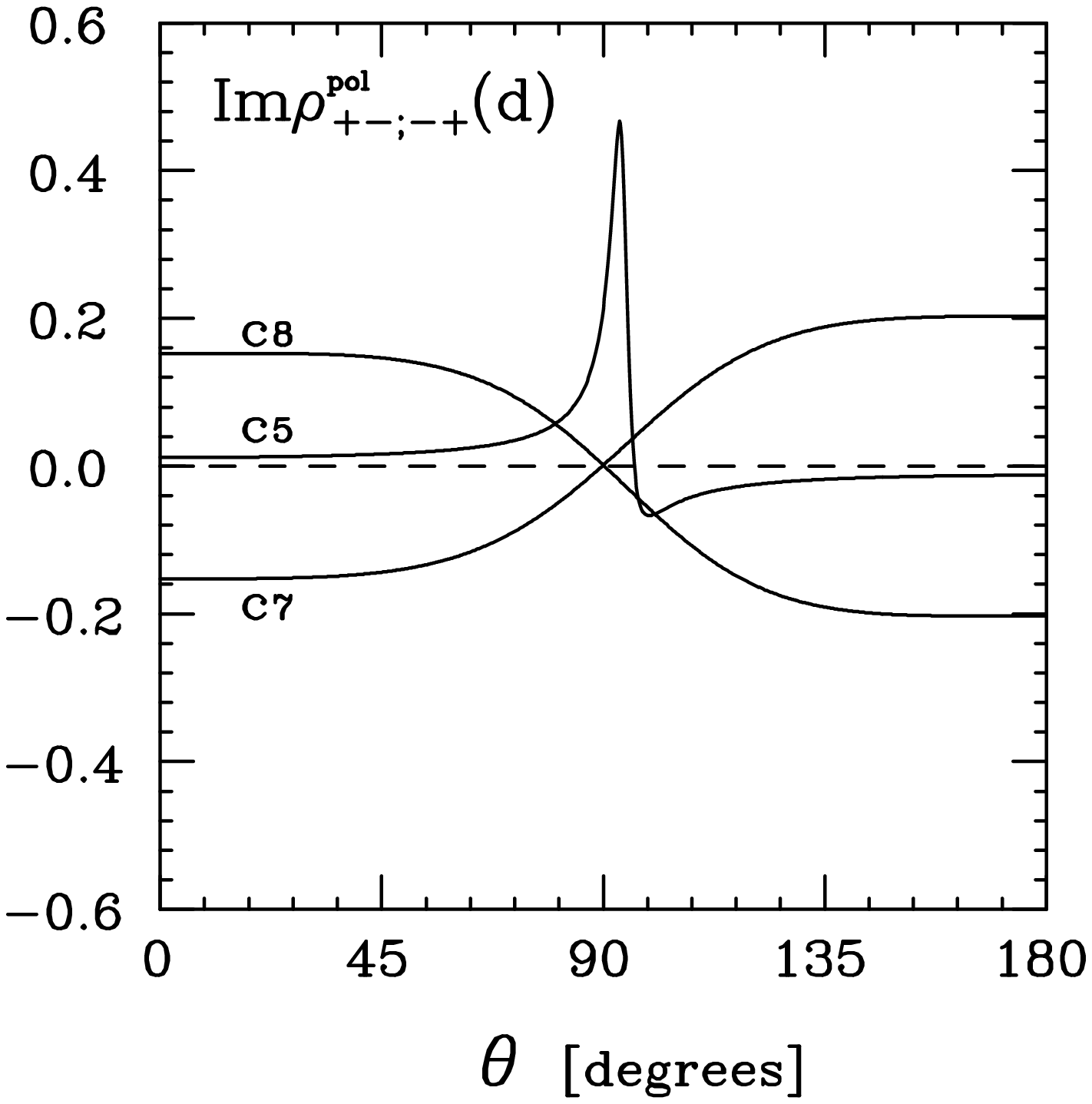,bbllx=50pt,bblly=200pt,bburx=530pt,%
bbury=650pt,width=15.0cm,height=15.0cm}}
 \begin{center}
 \vspace{12pt}
 \begin{minipage}[c]{13cm}
 {\small {\bf Fig. 4:}
The same as in Fig. 3, for $d$-type quarks. }
 \end{minipage}
 \end{center}
\end{figure}

\clearpage

\begin{figure}[c]
\centerline{
\epsfig{figure=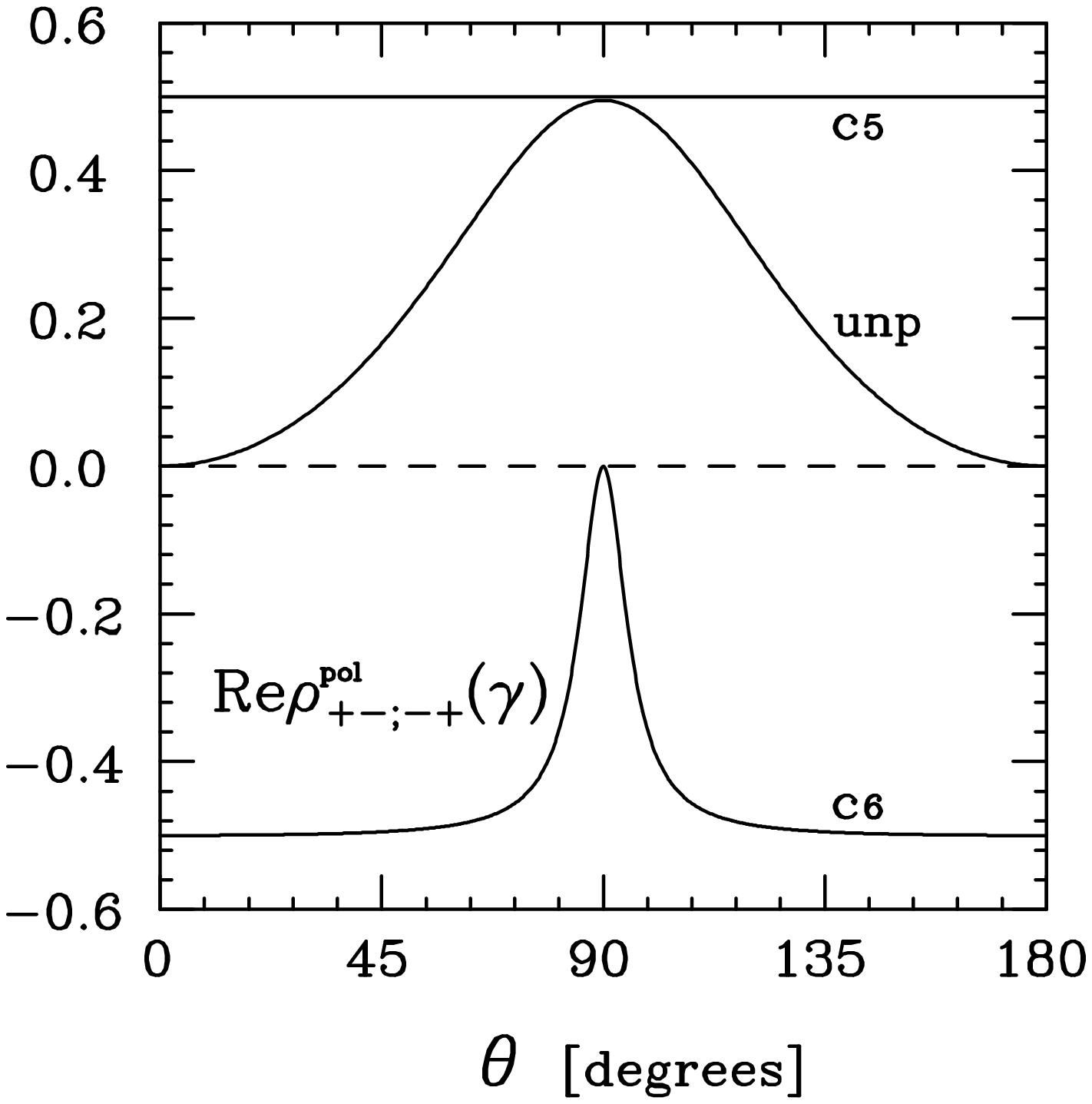,bbllx=50pt,bblly=200pt,bburx=530pt,%
bbury=650pt,width=15.0cm,height=15.0cm}}
 \begin{center}
 \vspace{12pt}
 \begin{minipage}[c]{13cm}
 {\small {\bf Fig. 5:}
Plot of Re[$\rho^{pol}_{+-;-+}(\qq; \sqrt s \ll M_{_Z})$] 
as a function of $\theta$ (the production angle of the vector meson in 
the $e^-e^+$ c.m. frame) for cases $C5$ and $C6$ 
(both leptons transversely polarized with 
spins either parallel or anti-parallel). All other cases give the same result 
given by unpolarized leptons, which is shown for comparison. Quark masses 
have been taken into account, with $\epsilon = 2m_q/\sqrt s = 0.1$. }
 \end{minipage}
 \end{center}
\end{figure}

\clearpage

\begin{figure}[c]
\centerline{
\epsfig{figure=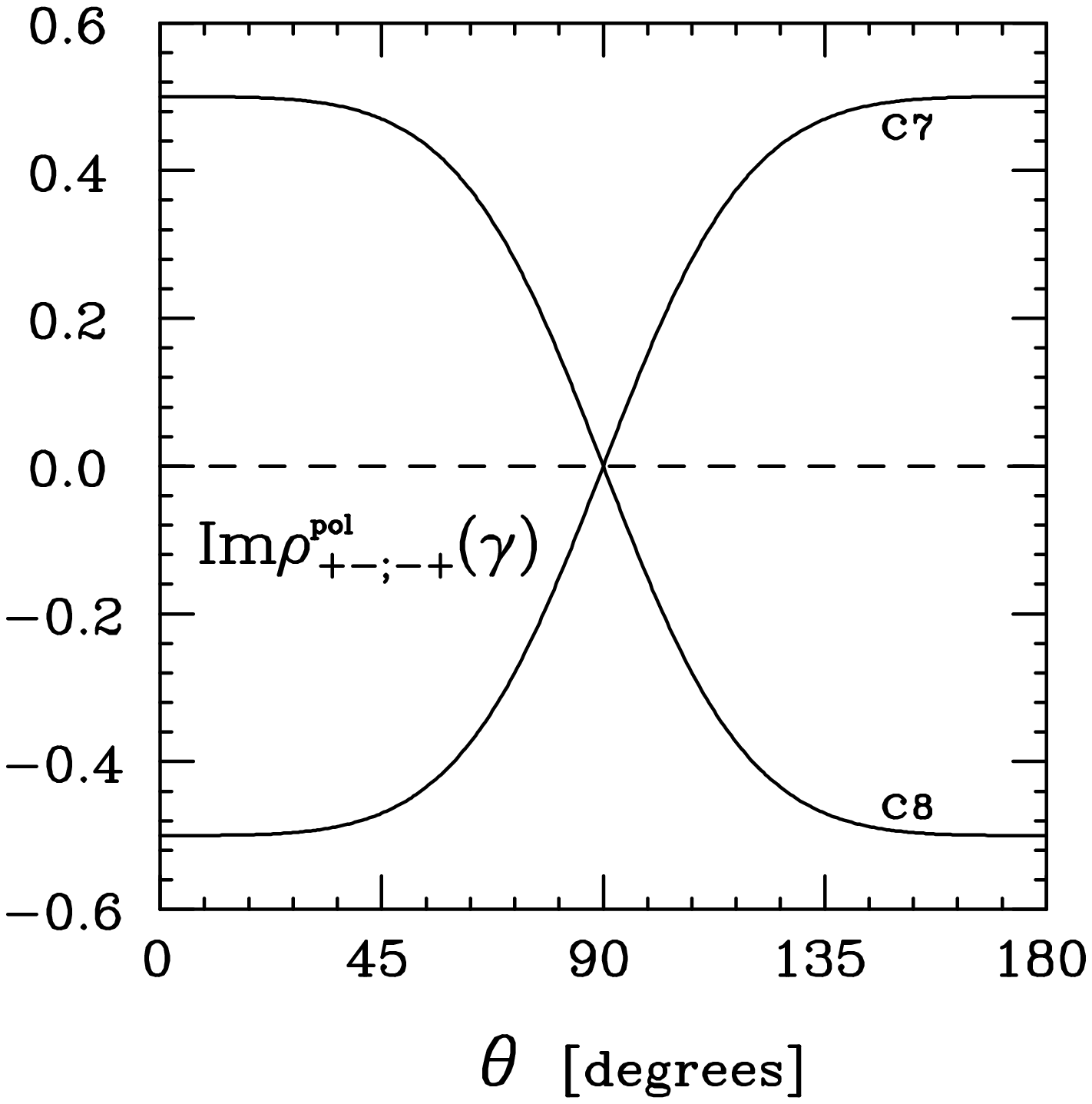,bbllx=50pt,bblly=200pt,bburx=530pt,%
bbury=650pt,width=15.0cm,height=15.0cm}}
 \begin{center}
 \vspace{12pt}
 \begin{minipage}[c]{13cm}
 {\small {\bf Fig. 6:}
Plot of Im[$\rho^{pol}_{+-;-+}(\qq; \sqrt s \ll M_{_Z})$] 
for cases $C7$ and $C8$ (both leptons transversely polarized, in
different directions). In all other cases, {\it including the unpolarized 
one}, Im[$\rho^{pol}_{+-;-+}(\qq; \sqrt s \ll M_{_Z})] = 0$.
Again, $\epsilon = 0.1$. }
 \end{minipage}
 \end{center}
\end{figure}

\end{document}